\documentclass[twocolumn,aps,prl]{revtex4}

\usepackage{graphicx}
\usepackage{dcolumn}
\usepackage{bm}
\usepackage{color}
\usepackage{hyperref} 

\begin{document}


\title{Optical spatial solitons in soft-matter: mode coupling theory approach}
\author{Claudio Conti,$^{1,2}$ Giancarlo Ruocco,$^{2}$ Stefano Trillo$^{3}$}
\email{claudio.conti@phys.uniroma1.it}
\affiliation{
$^{1}$Centro Studi e Ricerche ``Enrico Fermi'', Via Panisperna 89/A, 
00184 Rome, Italy \\
$^{2}$Research center SOFT INFM-CNR Universita' di Roma ``La Sapienza'',  P. A. Moro 
2, 00185, Roma, Italy\\
$^{3}$Dept.~of Engineering, University of Ferrara, Via Saragat 1, 
44100 Ferrara, Italy
}
\date{\today}  
\begin{abstract}
We predict that spatial self-trapping of light can occur in soft 
matter encompassing
a wide class of new materials such as colloids, foams, gels, fractal 
aggregates etc.
We develop a general nonlocal theory that allows
to relate the properties of the trapped state of Maxwell equations
to the measurable static structure factor of the specific material.
We give numerical evidence for stable trapping in fractal aggregates and
suggest also the possibility of soliton spectroscopy of soft-matter.
\end{abstract}

\maketitle Self-trapping of light beams, predicted forty years ago
\cite{Chiao64}, is still a subject of great interest 
\cite{TrilloBook,KivsharBook}. 
Observation of optical spatial
solitons (OSS) at low (down to mW) power levels, demonstrated in
photorefractive or liquid crystals,
makes OSS attractive candidates in several applications of
emerging photonics technology, 
\cite{Peccianti04nature,BoardmanBook,Stegeman00},
 and have driven successful 
efforts to understanding the role played by
specific material properties such as anisotropy and/or nonlocality
in self-trapping mechanisms. Yet, the description of light
trapping seems still strongly material dependent, and, as such,
cannot be applied to a whole class of condensed matter, namely
{\em soft matter} (SM), which encompasses the important case of
bio-matter where OSS can find new applications (e.g., laser
surgery, optical manipulation of nano-particles). Roughly
speaking, softness is generally due to a characteristic mesoscopic
(i.e., larger than atomic) length scale of the constituents, and
an energy scale comparable to room temperature thermal energy. 
As a 
consequence, SM properties can be easily tailored via external
field of different (mechanical, electrical, magnetic, thermal,
...) origin. SM includes colloidal suspensions, emulsions, and
foams (all involving different constituents in a host fluid), and
typical examples are polymers in a liquid, glues, liposomes,
blood, and all sort of bio-matter \cite{Likos01}.

Historically, the use of aerosols and water suspension of
dielectric spheres as nonlinear media dates back to the early 80's
\cite{Palmer80,Ashkin82}. However, in these materials
electrostriction has been described in the framework of simple
models that treat the diluted constituents as a gas of
non-interacting particles, in terms of {\em local} index change of
the Kerr type ($\Delta n=n_2 I$ where $I$ is the local intensity)
\cite{Chiao64,Palmer80,Ashkin82}. As a consequence existence of
stable OSS in two transverse dimensions is ruled out by the
occurrence of the well-known catastrophic self-focusing
instability \cite{Berge98}. Phenomenologically, stabilization can
be expected from the index saturation arising from the maximum
packing fraction of the dispersed particles. However, before
reaching such a regime, the physics of these materials, and in
general of other SM systems, is well-known to be affected by
particle-particle interactions. In particular, this occurs
whenever the particle-particle correlation function $g(r)$ is structured
on a length scale comparable to the laser beam waist. Under these
conditions a {\em nonlocal} model for self-focusing in SM must be
considered.

In this letter, we propose a novel general approach to stationary
self-focusing, linking for the first time the electrostrictive
nonlocal response of SM to its static structure factor
$S(q)$ (roughly speaking, the Fourier transform of the
particle-particle correlation function), usually measured by means
of scattering experiments. This allows us (i) to predict
stable propagation of two-dimensional OSS in a new wide class of
condensed matter, and (ii) to assess the importance that
ultra-focused laser light can have to investigate the properties
of SM. Both issues are of paramount importance in order to go
towards a more general description of solitons in complex media
(fractal aggregates, structured and supercooled liquid, etc.)
and their application in bio-photonics as well as to develop
a new spectroscopic tool for the investigation of SM properties.\\
\indent Assuming a linearly polarized beam and exploiting isotropy of system,
we start from the unidirectional scalar wave equation written for
a monochromatic beam with complex amplitude
$\mathcal{E}(x,y,z)$ propagating along $z$  \cite{Kolesik04}
\begin{equation} \label{unidir}
\displaystyle{i \frac{\partial \mathcal{E}}{\partial z} +
\sqrt{k^2 + \nabla_{\perp}^2} \mathcal{E} +
\frac{\omega}{2 c n_0} P_{nl}(\mathcal{E})
=0}
  \text{.}
\end{equation}
where $k^2=\omega^2 \epsilon(\omega)/c^2=\omega^2 n_0^2/c^2$, $n_0$ is the SM
bulk refractive index, and
$\nabla^2_{\perp}=\partial_x^2+\partial_y^2$. 
We further assume that the nonlinear polarization is responsible for a refractive
index change $\Delta n$, i. e. $P_{nl}(\mathcal{E})=\Delta \chi \mathcal{E}=2 n_{0}
\Delta n \mathcal{E}$, which is dominated by an electrostrictive
contribution (fast electronic nonlinearities as well as index
change due to thermal heating by optical absorption are usually
negligible, 
and we also neglect scattering). This legitimates our scalar (polarization-indipendent)
approach \cite{BoydBook}, regardless of beam spectral content
[at variance with Kerr effect of electronic origin where the non-paraxial regime 
at very high intensities requires to account for vectorial effects
(see e.g. \cite{Ciattoni02})].

Since the electrostriction $\Delta n= \rho (\partial
n/\partial \rho)_{\rho_0}$ is proportional to the particle number density
change $\rho$ (from equilibrium value $\rho_0$), Eq.
(\ref{unidir}) must be coupled to an evolution equation for
$\rho$. A widely accepted and largely applicable theory for SM is
the so-called Mode-Coupling Theory (MCT)
\cite{Bengtzelius84,Gotze99,Cummins99} which relies on the
so-called Zwangig-Mori formalism (given some observable, like
$\rho$, it allows to write closed equations for it and for its
correlations functions \cite{ZwanzigBook}). 
By exploiting MCT (see also \cite{Zaccarelli01}), we find that the density perturbation $\tilde
\rho=\tilde \rho(q,t)$ in Fourier space $q=(q_{x}, q_{y}, q_{z})$,
 obeys the dynamical
equation:
\begin{equation} \label{CMTequation}
\begin{array}{l}
\ddot{ \tilde \rho}(q,t)+q^2 \frac{k_B T}{S(q)}\tilde \rho+q^2
\int_0^t m(t-t') \dot{\tilde\rho}(q,t')dt'=\\
  \tilde f(q,t) + \frac{1}{2}\gamma_e \eta q^2  \tilde I(q)\text{.}
\end{array}
\end{equation}
where tilde denotes 3D spatial Fourier transform, 
$\eta=(\mu_0/\epsilon_0 n_0^2)^{1/2}$ is the 
impedance, $k_B$ is the Boltzman constant, $T$ is the temperature, $m(t)$ is the memory
function of the system (for a simple liquid $m(t)$ is a Dirac
delta function times the viscosity), and $S(q)$ is the static
structure factor. In Eq.~(\ref{CMTequation}) $f$ is a Langevin
term describing random forces (see e.g. \cite{Zaccarelli01}).

Eq. (\ref{CMTequation}) has been always considered without the deterministic forcing 
term weighted by the electrostrictive coefficient $\gamma_e=\rho (\partial
\epsilon/\partial \rho)_{\rho_0}$ 
\footnote{
Specific expressions for $\gamma_{e}$ 
follow from the dependence of the dielectric constant on particle density
$\epsilon(\rho)$, e.g. for a suspension of dielectric spheres of radius $r_s$, 
$\gamma_{e}=\rho_{0}
4 \pi r_s^{3} \epsilon_{h} (\epsilon_{s}-\epsilon_{h}
)/(\epsilon_{s}+2\epsilon_{h} )$, where subscripts $h$ and $s$
refer to the host medium and the spheres, respectively. 
}
, and provides one the most successful approaches to structural phase transitions 
of soft-matter. Here we extend it by accounting for the presence of an external optical field 
[coupled through Eq. (\ref{unidir})], which induces an electrostrictive force
with potential proportional (in configurational space) to  $\nabla^2 I $. 
This model can be thought of as a generalization of the acoustic wave equation
\cite{BoydBook}, which has been previously employed to determine the electrostrictive correction
to electronic nonlinearity of silica glass \cite{Buckland96}. 
The latter case is retrieved in our approach for $S(q)=\text{constant}$ and a memory-less response 
[i.e. $m(t)\propto\delta(t)$]. In the general case considered here, 
MCT accounts for the elastic deformation of a medium made of
interacting particles which results in a non-homogeneous response weighted by $S(q)$.
Additionally, though here we deal with electrostriction,
MCT is a powerful approach that can be generalized to account for other
types of nonlinearity, e.g. reorentational mechanisms, by looking
at different observables. As such, it provides a general framework for 
studying nonlinear optics in SM, going beyond the idealized Kerr (local) limit \cite{Palmer80,Ashkin82}
(which is, nevertheless, correctly retrieved for non-interacting particles, as shown below).
In the following, we address specifically the properties of spatial solitons.

Assuming that the random fluctuations are negligible 
with respect to the
driving electrostrictive term, the stationary state solution
($\partial_t=0$) of Eq. (\ref{CMTequation}) yields
\begin{equation} \label{rhosq}
\tilde \rho(q)=\frac{\gamma_e \eta}{2} \frac{S(q)}{k_B T}  \tilde I(q),
\end{equation}
which shows that $S(q)$ plays the role of a transfer function from
the optical intensity to the density. 
Incidentally, Eq. (\ref{rhosq}) can be also obtained by starting from a different model
of SM employing so-called generalized hydrodynamics equations
\cite{degrootbook} such as those typically adopted to modelling
inelastic light scattering spectra (ISTS) \cite{BernePecorabook}.
Obviously, in real space, Eq.~(\ref{rhosq}) corresponds to a
differential equation for $\rho=\rho(x,y,z)$. While this equation,
as well as Eq.~(\ref{unidir}), involves three dimensions, OSS
imply by definition a $z$-independent intensity. Therefore, in
this case, $\rho$ does not depend on $z$ and Eq. (\ref{rhosq}),
coupled to Eq. (\ref{unidir}), will be interpreted henceforth as a
2D transverse equation with $q=(q_{x},q_{y})$
\footnote{The 2D approach is justified also for input beams that do
not match exactly the OSS profile, since changes in $z$ occur 
usually on a length scale much longer than the transverse 
dependence of $\rho$ that yields the trapped state.}.\\

Equations (\ref{unidir}-\ref{rhosq}) allow us to develop a general
nonlocal model for trapping in SM. First we consider the paraxial
(Fresnel) regime, which corresponds to expanding the transverse operator in
the $q$-Fourier transform of Eq. (\ref{unidir}) as $\sqrt{k^{2} -
q^2} \simeq k [1-q^2/(2k^2)]$. Equation (\ref{unidir}) becomes,
in 
terms of the slowly varying envelope $E(x,y,z)=\mathcal{E}(x,y,z) \exp(-i 
k z)$,
\begin{equation} \label{unidir2}
\displaystyle{2ik \frac{\partial E}{\partial z}
+
\nabla_{\perp}^2 E + \frac{2k^2}{n_0}\left(\frac{\partial n}{\partial 
\rho}\right)_{\!\!\rho_0}\!\!\! \rho E =0}
  \text{.}
\end{equation}
In Eq. (\ref{unidir2}) the nonlinearity arises from the term $\rho E$.
Indeed, back-transforming Eq. (\ref{rhosq}) to real
space for $\rho$ and assuming azimuthal symmetry, we obtain, 
after some algebra, 
a self-consistent nonlinear non-local wave equation
\begin{equation} \label{Fock}
i 2 k \displaystyle \frac{\partial E}{\partial z}
+
\nabla_{\perp}^2 E + \chi E \! \!\int_0^{\infty} \! \! \! \!
G(r,r') |E(r',z)|^2 \; r' dr'=0\text{,}
\end{equation}
where $r \equiv \sqrt{x^2+y^2}$, and we have defined the kernel
\begin{equation} \label{kernel}
G(r,r') \equiv \int_0^{\infty}
\frac{S(Q)}{S_{0}} J_0(Q r) J_0(Q r')~Q dQ,
\end{equation}
and the coefficient
$\chi \equiv k^2 (\partial n/\partial \rho)_{\rho_0} \gamma_e S_{0}/2 
k_B T n_0$
(note that $S(q)$ is scaled to $S_0=S(0)$ representing the ratio between the
compressibility of the material and that of the ideal gas
\cite{SimpleLiquids}). We seek for bound states
of Eqs. (\ref{Fock}-\ref{kernel}) in the form $E(z, r)= (\chi w_{0})^{-1/2}
u(\sigma) \exp(i \beta \zeta)$, where $\beta$ is the nonlinear correction to the wavevector $k$, 
which is determined self-consistenly in the numerical simulations,
and $\sigma=r/w_0$, $\zeta=z/z_0=z/2 k w_0^2$ 
are 
dimensionless radial and longitudinal variables,
in units of beam 
width $w_0$ and diffraction (Rayleigh) length $z_0$, respectively.
The OSS (bound state) profile $u(\sigma)$ obeys the non-local 
eigenvalue equation
(we set $w_{0}^{2} \nabla_{\perp}^{2} \equiv 
\nabla^2_{\sigma}=d^2/d\sigma^2 + \sigma^{-1} d/d\sigma$)
\begin{equation} \label{boundstates}
\nabla_\sigma^2 u - \beta u + u \int_0^{\infty}
g(\sigma,\sigma')u^2(\sigma')~\sigma' d\sigma'=0,
\end{equation}
where the kernel $g(\sigma, \sigma')$ can be obtained (at least 
numerically), once
$S(Q)$ is known, from the integral
\begin{equation} \label{g}
g(\sigma,\sigma')=\int_{0}^{\infty}
\frac{ S(\theta/w_0)}{S_0} J_0(\sigma' \theta) J_0(\sigma 
\theta)~\theta d\theta \text{.}
\end{equation}
 From Eqs. (\ref{boundstates}-\ref{g}) the {\em ideal local} (Kerr) limit
is recovered for $S(q)=S_{0}=constant$, which yields
$g(\sigma,\sigma')= \delta(\sigma-\sigma')/\sigma$, 
and hence a pure 
Kerr law with a nonlinear index that, 
in the case of the dielectric 
spheres, turns out to be
\begin{equation}
\label{n2sq}
n_{2}=\frac{4\pi^2 r_s^6 n_h^4}{c n_0^2}\left(
\frac{\epsilon_s-\epsilon_h}{\epsilon_s+2\epsilon_h}\right)^2
\frac{\rho_0 S_0}{k_B T}.
\end{equation}
This limit, however, is well known to lead to unstable 
(so-called Townes after Ref. \cite{Chiao64}) OSS. 
Conversely, in 
the general case $S(q) \neq \text{constant}$,
we expect solutions of Eqs. 
(\ref{boundstates}-\ref{g})  
to be stabilized by non-locality 
\cite{Bang02,Krolikowski04,Briedis05,Yakimenko05,Rotschild05}.
We are also naturally brought to consider 
deviation
from paraxiality due to strong focusing, 
and argue for the 
existence of OSS in this
case. In fact, narrow OSS in SM may be important both in specific
applications (e.g. laser surgery) and in order to establish
OSS as a 
mean for probing the static structure factor
$S(q)$ of SM when the latter extends to high spatial frequencies.
Deviations from paraxiality can be accounted for by considering
the next order in the expansion of the transverse operator in Eq.
(\ref{unidir}). By adopting the normalization employed for the
paraxial case, we cast the new bound state equation in the form
\begin{equation} \label{boundstatesnonparaxial}
\nabla_\sigma^2 u-\varepsilon \nabla_\sigma^4 u -\beta u+
u \int_0^{\infty} g(\sigma,\sigma')u^2(\sigma')~\sigma' d\sigma' = 0\text{,}
\end{equation}
where the degree of non-paraxialiy is measured by a single
dimensionless parameter $\varepsilon=(\lambda/4\pi n w_0)^2$ fixed
by the ratio between wavelength $\lambda$ and beam width scale $w_{0}$.

In order to discuss various OSS supported by different types of
SM, we make specific examples. We start considering hard spheres
in a host liquid (solvent). In the limit of diluted, non
interacting spheres, $S(q)$ is constant and this yields, once
again, unstable OSS. In a more refined approximation, $S(q)$ can
be described by a parabolic law in the framework of the
Percus-Yevick model \cite{SimpleLiquids}
\begin{equation} \label{quadraticSq}
S(q)= S_0 + K q^2.
\end{equation}
After Eq. (\ref{rhosq}), the corresponding expression for $\rho$ (as stated, we assume $\rho$ to follow adiabatically $I$ along $z$) is
\begin{equation} \label{xyparabolic}
\rho(r,z)=\frac{\gamma_e \eta}{2 k_B T}\left[S_0 I(r,z) - K \nabla_\perp^2 I(r,z)\right]
\end{equation}
which, once inserted in Eq.~(\ref{unidir2}), gives a model for {\em weakly non-local} solitons that has
interdisciplinary interest (plasma physics, matter waves,
transport in DNA, see Ref.~\cite{Bang02} and references therein).
Stable soliton solutions of this model have been reported
\cite{Bang02} and, in this context, represent 1+2D OSS in SM, when
its static structure factor can be well approximated by Eq.
(\ref{quadraticSq}). To this end, consider that, using
Percus-Yevick model, the parabolic approximation of $S(q)$ breaks
down around $q r_{s} \cong 5$. Since $q$ can be reasonably
estimated to be $q \sim w_{0}^{-1}$, the weakly non-local model
starts to loose its validity when the spheres have size comparable
with beam width $w_{0}$ (in this regime, a microscopic
description of the molecular dynamics is needed). Viceversa, when
Eqs. (\ref{quadraticSq}-\ref{xyparabolic}) hold valid, the
nonlocality that, generally speaking, provides the stabilizing
mechanism of OSS against catastrophic self-focusing \cite{Bang02}
stems from the particle-particle correlation function $g(r)$,
which is proportional to the Fourier transform of $S(q)-1$
\cite{SimpleLiquids}, as anticipated. Importantly, since
$S(q)$ is not uniform,  stable OSS not only exist, but provide
information on the material [for a given optical power,
the width of OSS is determined by the constant $K$
in Eq.~(\ref{quadraticSq})].\\
\indent Further models for $S(q)$ can be discussed. A very
intriguing case is that in which the suspended particles of
colloidal SM develop self-similar aggregates with fractal
dimension $D$, described by the function \cite{FractalBook}
\begin{equation} \label{fractalSQ}
\begin{array}{l}
S(q)=1\!+\! \frac{D~\Gamma(D-1)}{(q r_s)^D}
\left(\frac{q^2 \xi^2+1}{q^2 
\xi^2}\right)^{\frac{1-D}{2}}\!\sin[(D-1) \tan^{-1}(q \xi)]
\end{array}
\end{equation}
where $\Gamma$ is the Gamma function, $r_s$ is the spheres radius,
and $\xi$ gives the spatial extension of the aggregate.
Incidentally, when $D=2$, Eq. (\ref{fractalSQ}) yields
$S(q)=1+2 \left(\xi/r_s \right)^2 \left(1 + q^2
\xi^2\right)^{-1}$, which entails the sum of a Kerr contribution
and a non-local one with Lorentzian lineshape.
In the limit $q \xi << 1$, Eq. (\ref{fractalSQ}) yields
\begin{equation}
S(q)=\Gamma(D+1) \frac{\xi}{r_s} \left[1-\frac{D(D+1)}{6} q^2 \xi^2 
\right] \text{,}
\end{equation}
and the corresponding model for $\rho$ reads as:
\begin{equation} \label{fractalro}
\left[1-\frac{D(D+1)}{6}\xi^2 \nabla_\perp^2 \right] \rho=
\frac{\eta \gamma_e \Gamma(D+1)(\xi/r_s)}{2 k_{B} T} I\text{.}
\end{equation}
 From Eq.~(\ref{fractalro}) it is readily seen that $\rho$
spatially decays (when $I=0$) as the modified Bessel function
$K_0$ of argument $\xi \sqrt{D(D+1)/6} \sigma$
which depends on the fractal dimension. Recalling that $\sigma=r/w_{0}$ and that the degree of nonlocality is 
the ratio between the spatial decay rate of the
optically induced index perturbation and that of the self-trapped
beams, our result implies that \textit{the degree of optical
non-locality scales basically as the fractal dimension of the
material}. Noteworthy, Eqs. (\ref{Fock})-(\ref{fractalro}) define
another well known model for non-local OSS, which applies in the
case of nematic liquid crystals \cite{Conti03}.

\begin{figure}
\includegraphics[width=8.3cm]{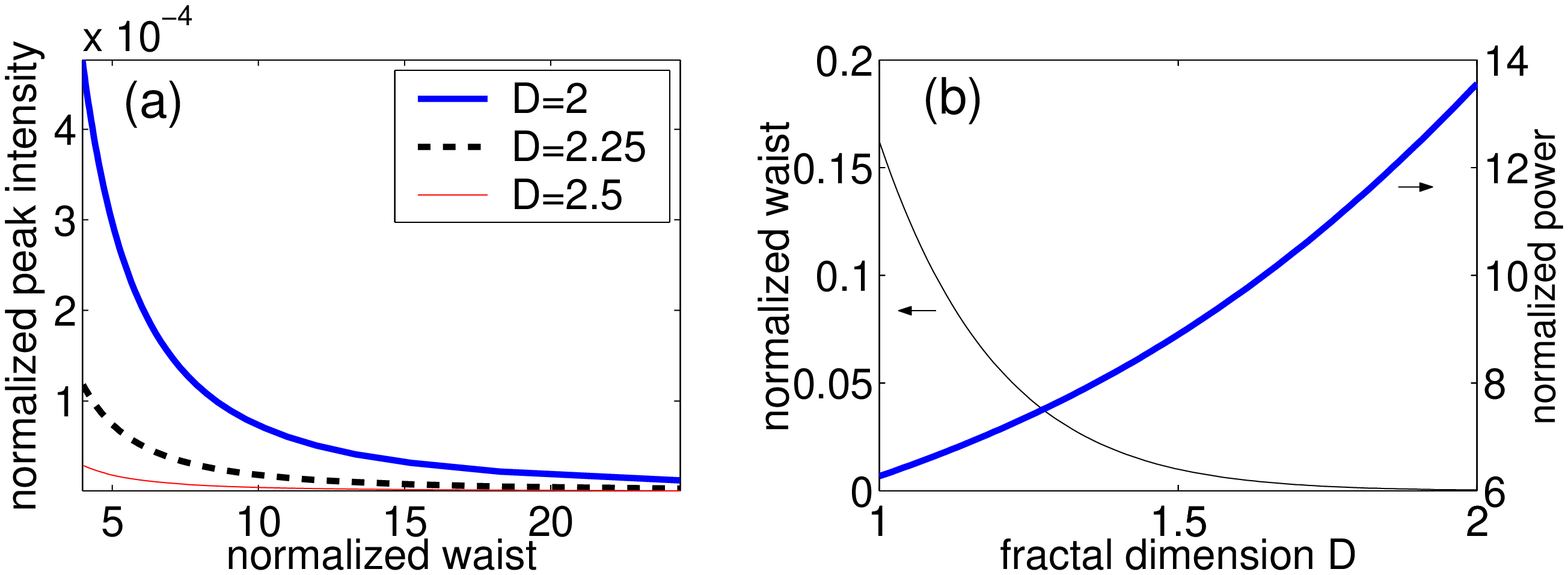}
\caption{
 (a)Normalized peak intensity $max(u^2)$ Vs. the 
normalized beam width
(i.e. the std of $u^2$), obtained from Eqs. (\ref{boundstates}-\ref{g})
coupled to Eq. (\ref{fractalSQ}) for various fractal dimensions $D$.
(b) Normalized beam waist and power Vs. the fractal
dimension $D$ for $\beta=2$.
\label{figure1}}
\end{figure}

\begin{figure}
\includegraphics[width=8.3cm]{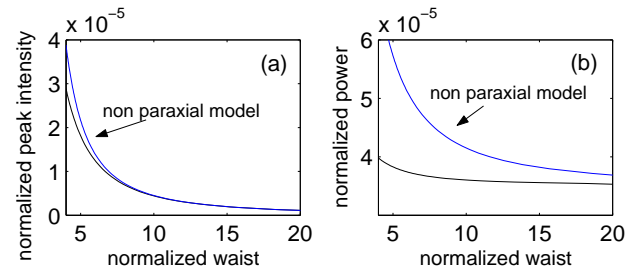}
\caption{  As in figure \ref{figure1}, soliton normalized peak intensity (a) and power (b)
Vs. normalized waist for the paraxial ($\epsilon=0$) and non paraxial
($\epsilon=0.1$) model ($D=2.5$, $\xi=100r_s$, $\kappa=0.1$).
  \label{figure2}}
\end{figure}

In order to show that OSS exist also in the general case 
[Eq.~(\ref{fractalSQ})], with features directly linked to the fractal dimension $D$,
we resort to numerical integration of Eqs. (\ref{boundstates}-\ref{g}) [or
Eqs. (\ref{boundstatesnonparaxial}-\ref{g}) in the non-paraxial regime]
using finite difference discretization in $\sigma$ and Newton-Rapson 
iterations.
To fix the ideas, we show results for the
characteristic values of the following length scale ratios
(between aggregate dimension $\xi$, particle radius $r_{s}$, beam width $w_0$):
$\xi/r_s=100$ and $\kappa=\xi/w_0=0.1$.
In Fig.~\ref{figure1} (a) we show {\it existence curves}, 
i.e. the 
soliton normalized peak intensity against the normalized
soliton width [$max(u^2)$ vs. {\it std of $u^2$} parametrized by $\beta$], obtained for three values of $D$.
As shown the features of OSS change with fractal dimension $D$.
This is even more clear from Fig.~\ref{figure1} (b),
where we display the OSS normalized power (i.e. the norm $Q=2\pi\int u^2 r dr$)
and width as a function of the fractal dimension $D$
(here we fix $\beta=2$).
In Fig.~ \ref{figure2} we show the effect of non-paraxiality.
The features of OSS starts to exhibit significant deviations when the
soliton width decreases significantly, and non-paraxial effects are 
no longer negligible.\\

It is interesting to observe that for a fractal medium the equation for $\rho$ in the configurational space
includes fractional derivatives. Soft matter, not only provide a un-precedented framework to study nonlocality
with a taylorable structure factor, but also opens the may to new mathematical models, as we will discuss in future
publications.

\begin{figure}
\includegraphics[width=8.3cm]{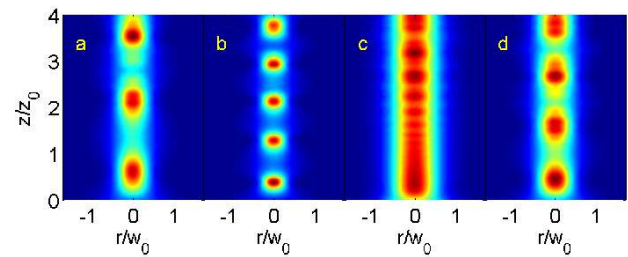}
\caption{
Propagation (symmetrized for $r<0$) of an input Gaussian beam 
$u(\zeta=0,r)=A \exp(-\sigma^2)$
in the paraxial regime ($\epsilon=0$): (a) $A=0.0055$; (b) $A=0.006$; 
and 
beyond  the paraxial regime ($\epsilon=0.01$); (c) $A=0.0055$; 
(d) $A=0.006$. Stable trapping when $\epsilon=0$ is achieved for $A=0.0052$
($D=2.5$, $\kappa=0.1$, $\xi=100r_s$).
}
\label{figure3}
\end{figure}

\indent To prove that stable self-trapping can be 
achieved for input conditions that do not exactly match the OSS 
profile, we have run beam propagation simulations for the paraxial 
and non-paraxial models. 
In Fig.~ \ref{figure3} we show the spatial evolution of an input
Gaussian TEM$_{00}$ laser beam $u(\sigma, \zeta=0)=A\exp(-\sigma^{2})$, whose parameters do not exactly match
the existence condition. 
The resulting longitudinal beam 
oscillations, which depend on beam power
(for a fixed width) and 
degree of non-paraxiality, have been reported
previously for other non-local solitons \cite{Conti04}. 
They are connected with the fact that non-local
solitons are absolutely stable, \cite{Bang02}
and can be related to excitation of  ``internal modes'' of
the soliton \cite{KivsharBook,Rosanov2002}, or, in the framework of the hyghly 
non-local approximation
[i.e. when  $I(q)$ varies on a q-scale much broader
than that of $S(q)$]
to the existence of exact breathing solutions 
\cite{Snyder97,Conti04,Guo04}. 
Notably, in the latter regime, $\tilde I(q) \simeq \tilde I(0) 
\equiv  P$ in Eq. (\ref{rhosq}), with $P$ the optical beam power. 
In this case, Eq. (\ref{unidir2}) takes the form of the linear Schr\"{o}dinger equation 
for a quantum particle in a 2D potential well with shape dictated 
by particle autocorrelation $g(r)$ and power $P$,
and OSS reduce to the bound states, which can be found by standard 
techniques.
Such results let us envisage a broad physical setting for the observation
of deeply oscillating nonlocal solitons and show that their 
existence is not restricted to the paraxial regime,

In summary we have shown that optical beams can be self-trapped
in soft matter both in the paraxial and tightly focusing regimes,
thus opening new perspectives for their applications in biological materials
and as a mean to probe properties of condensed matter.
For example, fixing the input beam waist and adjusting the incoming
power to find the appearance of a soliton allows one to directly measure the fractal dimension of
the aggregates.
Assessing the role of time dynamics and thermal contributions
as well as the validity of the present approach in other
condensed matter systems (e.g., supercooled liquids, where optical 
trapping is unexplored to date)
will be natural extensions of this work.

We acknowledge fruitful discusssions with F. Sciortino, and funding 
from MIUR (PRIN
and FIRB projects).


\begin{thebibliography}{33}
\expandafter\ifx\csname natexlab\endcsname\relax\def\natexlab#1{#1}\fi
\expandafter\ifx\csname bibnamefont\endcsname\relax
  \def\bibnamefont#1{#1}\fi
\expandafter\ifx\csname bibfnamefont\endcsname\relax
  \def\bibfnamefont#1{#1}\fi
\expandafter\ifx\csname citenamefont\endcsname\relax
  \def\citenamefont#1{#1}\fi
\expandafter\ifx\csname url\endcsname\relax
  \def\url#1{\texttt{#1}}\fi
\expandafter\ifx\csname urlprefix\endcsname\relax\def\urlprefix{URL }\fi
\providecommand{\bibinfo}[2]{#2}
\providecommand{\eprint}[2][]{\url{#2}}

\bibitem[{\citenamefont{Chiao et~al.}(1964)\citenamefont{Chiao, Garmire, and
  Townes}}]{Chiao64}
\bibinfo{author}{\bibfnamefont{R.~Y.} \bibnamefont{Chiao}},
  \bibinfo{author}{\bibfnamefont{E.}~\bibnamefont{Garmire}}, \bibnamefont{and}
  \bibinfo{author}{\bibfnamefont{C.~H.} \bibnamefont{Townes}},
  \bibinfo{journal}{\prl} \textbf{\bibinfo{volume}{13}}, \bibinfo{pages}{479}
  (\bibinfo{year}{1964}).

\bibitem[{\citenamefont{Trillo and Torruealls}(2001)}]{TrilloBook}
\bibinfo{editor}{\bibfnamefont{S.}~\bibnamefont{Trillo}} \bibnamefont{and}
  \bibinfo{editor}{\bibfnamefont{W.}~\bibnamefont{Torruealls}}, eds.,
  \emph{\bibinfo{title}{Spatial solitons}}
  (\bibinfo{publisher}{Springer-Verlag}, \bibinfo{address}{Berlin},
  \bibinfo{year}{2001}).

\bibitem[{\citenamefont{Kivshar and Agrawal}(2003)}]{KivsharBook}
\bibinfo{author}{\bibfnamefont{Y.}~\bibnamefont{Kivshar}} \bibnamefont{and}
  \bibinfo{author}{\bibfnamefont{G.~P.} \bibnamefont{Agrawal}},
  \emph{\bibinfo{title}{Optical solitons}} (\bibinfo{publisher}{Academic
  Press}, \bibinfo{address}{New York}, \bibinfo{year}{2003}).

\bibitem[{\citenamefont{Peccianti et~al.}(2004)\citenamefont{Peccianti, Conti,
  Assanto, De~Luca, and Umeton}}]{Peccianti04nature}
\bibinfo{author}{\bibfnamefont{M.}~\bibnamefont{Peccianti}},
  \bibinfo{author}{\bibfnamefont{C.}~\bibnamefont{Conti}},
  \bibinfo{author}{\bibfnamefont{G.}~\bibnamefont{Assanto}},
  \bibinfo{author}{\bibfnamefont{A.}~\bibnamefont{De~Luca}}, \bibnamefont{and}
  \bibinfo{author}{\bibfnamefont{C.}~\bibnamefont{Umeton}},
  \bibinfo{journal}{Nature} \textbf{\bibinfo{volume}{432}},
  \bibinfo{pages}{733} (\bibinfo{year}{2004}).

\bibitem[{\citenamefont{Boardman and Sukhorukov}(2001)}]{BoardmanBook}
\bibinfo{editor}{\bibfnamefont{A.~D.} \bibnamefont{Boardman}} \bibnamefont{and}
  \bibinfo{editor}{\bibfnamefont{A.~P.} \bibnamefont{Sukhorukov}}, eds.,
  \emph{\bibinfo{title}{Soliton Driven Photonics}} (\bibinfo{publisher}{Kluwer
  Academic Publ.}, \bibinfo{address}{Dordrecht}, \bibinfo{year}{2001}).

\bibitem[{\citenamefont{Stegeman et~al.}(2000)\citenamefont{Stegeman,
  Christodoulides, and Segev}}]{Stegeman00}
\bibinfo{author}{\bibfnamefont{G.~I.~A.} \bibnamefont{Stegeman}},
  \bibinfo{author}{\bibfnamefont{D.~N.} \bibnamefont{Christodoulides}},
  \bibnamefont{and} \bibinfo{author}{\bibfnamefont{M.}~\bibnamefont{Segev}},
  \textbf{\bibinfo{volume}{6}}, \bibinfo{pages}{1419} (\bibinfo{year}{2000}).

\bibitem[{\citenamefont{Likos}(2001)}]{Likos01}
\bibinfo{author}{\bibfnamefont{C.~N.} \bibnamefont{Likos}},
  \bibinfo{journal}{Physics Reports} \textbf{\bibinfo{volume}{348}},
  \bibinfo{pages}{267} (\bibinfo{year}{2001}).

\bibitem[{\citenamefont{Palmer}(1980)}]{Palmer80}
\bibinfo{author}{\bibfnamefont{A.~J.} \bibnamefont{Palmer}},
  \bibinfo{journal}{\ol} \textbf{\bibinfo{volume}{5}}, \bibinfo{pages}{54}
  (\bibinfo{year}{1980}).

\bibitem[{\citenamefont{Ashkin et~al.}(1982)\citenamefont{Ashkin, Dziedzic, and
  Smith}}]{Ashkin82}
\bibinfo{author}{\bibfnamefont{A.}~\bibnamefont{Ashkin}},
  \bibinfo{author}{\bibfnamefont{J.~M.} \bibnamefont{Dziedzic}},
  \bibnamefont{and} \bibinfo{author}{\bibfnamefont{P.~W.} \bibnamefont{Smith}},
  \bibinfo{journal}{\ol} \textbf{\bibinfo{volume}{7}}, \bibinfo{pages}{276}
  (\bibinfo{year}{1982}).

\bibitem[{\citenamefont{Berge}(1998)}]{Berge98}
\bibinfo{author}{\bibfnamefont{L.}~\bibnamefont{Berge}},
  \bibinfo{journal}{Phys. Rep.} \textbf{\bibinfo{volume}{303}},
  \bibinfo{pages}{259} (\bibinfo{year}{1998}).

\bibitem[{\citenamefont{Kolesik et~al.}(2004)\citenamefont{Kolesik, Wright, and
  Moloney}}]{Kolesik04}
\bibinfo{author}{\bibfnamefont{M.}~\bibnamefont{Kolesik}},
  \bibinfo{author}{\bibfnamefont{E.~M.} \bibnamefont{Wright}},
  \bibnamefont{and} \bibinfo{author}{\bibfnamefont{J.~V.}
  \bibnamefont{Moloney}}, \bibinfo{journal}{\prl}
  \textbf{\bibinfo{volume}{92}}, \bibinfo{pages}{253901}
  (\bibinfo{year}{2004}).

\bibitem[{\citenamefont{Boyd}(1992)}]{BoydBook}
\bibinfo{author}{\bibfnamefont{R.~W.} \bibnamefont{Boyd}},
  \emph{\bibinfo{title}{Nonlinear Optics}} (\bibinfo{publisher}{Academic
  Press}, \bibinfo{address}{San Diego}, \bibinfo{year}{1992}).

\bibitem[{\citenamefont{Ciattoni et~al.}(2002)\citenamefont{Ciattoni, Conti,
  DelRe, Di~Porto, Crosignani, and Yariv}}]{Ciattoni02}
\bibinfo{author}{\bibfnamefont{A.}~\bibnamefont{Ciattoni}},
  \bibinfo{author}{\bibfnamefont{C.}~\bibnamefont{Conti}},
  \bibinfo{author}{\bibfnamefont{E.}~\bibnamefont{DelRe}},
  \bibinfo{author}{\bibfnamefont{P.}~\bibnamefont{Di~Porto}},
  \bibinfo{author}{\bibfnamefont{B.}~\bibnamefont{Crosignani}},
  \bibnamefont{and} \bibinfo{author}{\bibfnamefont{A.}~\bibnamefont{Yariv}},
  \bibinfo{journal}{\ol} \textbf{\bibinfo{volume}{27}}, \bibinfo{pages}{734}
  (\bibinfo{year}{2002}).

\bibitem[{\citenamefont{Bengtzelius et~al.}(1984)\citenamefont{Bengtzelius,
  G\"otze, and Sj\"olander}}]{Bengtzelius84}
\bibinfo{author}{\bibfnamefont{U.}~\bibnamefont{Bengtzelius}},
  \bibinfo{author}{\bibfnamefont{W.}~\bibnamefont{G\"otze}}, \bibnamefont{and}
  \bibinfo{author}{\bibfnamefont{A.}~\bibnamefont{Sj\"olander}},
  \bibinfo{journal}{J. Phys. C.:Solid State Phys.}
  \textbf{\bibinfo{volume}{17}}, \bibinfo{pages}{5915} (\bibinfo{year}{1984}).

\bibitem[{\citenamefont{G\"otze}(1999)}]{Gotze99}
\bibinfo{author}{\bibfnamefont{W.}~\bibnamefont{G\"otze}}, \bibinfo{journal}{J.
  Phys. C} \textbf{\bibinfo{volume}{11}}, \bibinfo{pages}{A1}
  (\bibinfo{year}{1999}).

\bibitem[{\citenamefont{Cummins}(1999)}]{Cummins99}
\bibinfo{author}{\bibfnamefont{H.~Z.} \bibnamefont{Cummins}},
  \bibinfo{journal}{J. Phys. C.} \textbf{\bibinfo{volume}{11}},
  \bibinfo{pages}{A95} (\bibinfo{year}{1999}).

\bibitem[{\citenamefont{Zwanzig}(2001)}]{ZwanzigBook}
\bibinfo{author}{\bibfnamefont{R.}~\bibnamefont{Zwanzig}},
  \emph{\bibinfo{title}{Nonequilibrium Statistical Mechanics}}
  (\bibinfo{publisher}{Oxford University Press, New York},
  \bibinfo{year}{2001}).

\bibitem[{\citenamefont{Zaccarelli et~al.}(2001)\citenamefont{Zaccarelli,
  Foffi, Sciortino, Tartaglia, and Dawson}}]{Zaccarelli01}
\bibinfo{author}{\bibfnamefont{E.}~\bibnamefont{Zaccarelli}},
  \bibinfo{author}{\bibfnamefont{G.}~\bibnamefont{Foffi}},
  \bibinfo{author}{\bibfnamefont{F.}~\bibnamefont{Sciortino}},
  \bibinfo{author}{\bibfnamefont{P.}~\bibnamefont{Tartaglia}},
  \bibnamefont{and} \bibinfo{author}{\bibfnamefont{K.~A.}
  \bibnamefont{Dawson}}, \bibinfo{journal}{Europhysics Lett.}
  \textbf{\bibinfo{volume}{55}}, \bibinfo{pages}{157} (\bibinfo{year}{2001}),
  \bibinfo{note}{see also arXiv:cond-mat/0111033}.

\bibitem[{\citenamefont{Buckland and Boyd}(1996)}]{Buckland96}
\bibinfo{author}{\bibfnamefont{E.~L.} \bibnamefont{Buckland}} \bibnamefont{and}
  \bibinfo{author}{\bibfnamefont{R.~W.} \bibnamefont{Boyd}},
  \bibinfo{journal}{\ol} \textbf{\bibinfo{volume}{21}}, \bibinfo{pages}{1117}
  (\bibinfo{year}{1996}).

\bibitem[{\citenamefont{de~Groot and Mazur}(1984)}]{degrootbook}
\bibinfo{author}{\bibfnamefont{S.~R.} \bibnamefont{de~Groot}} \bibnamefont{and}
  \bibinfo{author}{\bibfnamefont{P.}~\bibnamefont{Mazur}},
  \emph{\bibinfo{title}{Non-equilibrium thermodynamics}}
  (\bibinfo{publisher}{Dover, New York}, \bibinfo{year}{1984}).

\bibitem[{\citenamefont{Berne and Pecora}(2000)}]{BernePecorabook}
\bibinfo{author}{\bibfnamefont{B.~J.} \bibnamefont{Berne}} \bibnamefont{and}
  \bibinfo{author}{\bibfnamefont{R.}~\bibnamefont{Pecora}},
  \emph{\bibinfo{title}{Dynamic Light Scattering: With Applications to
  Chemistry, Biology, and Physics}} (\bibinfo{publisher}{Dover, New York},
  \bibinfo{year}{2000}).

\bibitem[{\citenamefont{Hansen and McDonald}(1986)}]{SimpleLiquids}
\bibinfo{author}{\bibfnamefont{J.-P.} \bibnamefont{Hansen}} \bibnamefont{and}
  \bibinfo{author}{\bibfnamefont{I.~R.} \bibnamefont{McDonald}},
  \emph{\bibinfo{title}{Theory of simple liquids}}
  (\bibinfo{publisher}{Academic Press}, \bibinfo{address}{London, UK},
  \bibinfo{year}{1986}), \bibinfo{edition}{2nd} ed.

\bibitem[{\citenamefont{Bang et~al.}(2002)\citenamefont{Bang, Krolikowski,
  Wyller, and Rasmussen}}]{Bang02}
\bibinfo{author}{\bibfnamefont{O.}~\bibnamefont{Bang}},
  \bibinfo{author}{\bibfnamefont{W.}~\bibnamefont{Krolikowski}},
  \bibinfo{author}{\bibfnamefont{J.}~\bibnamefont{Wyller}}, \bibnamefont{and}
  \bibinfo{author}{\bibfnamefont{J.~J.} \bibnamefont{Rasmussen}},
  \bibinfo{journal}{\pre} \textbf{\bibinfo{volume}{66}},
  \bibinfo{pages}{046619} (\bibinfo{year}{2002}).

\bibitem[{\citenamefont{Krolikowski et~al.}(2004)\citenamefont{Krolikowski,
  Bang, Nikolov, Neshev, Wyller, Rasmussen, and Edmundson}}]{Krolikowski04}
\bibinfo{author}{\bibfnamefont{W.}~\bibnamefont{Krolikowski}},
  \bibinfo{author}{\bibfnamefont{O.}~\bibnamefont{Bang}},
  \bibinfo{author}{\bibfnamefont{N.~I.} \bibnamefont{Nikolov}},
  \bibinfo{author}{\bibfnamefont{D.}~\bibnamefont{Neshev}},
  \bibinfo{author}{\bibfnamefont{J.}~\bibnamefont{Wyller}},
  \bibinfo{author}{\bibfnamefont{J.~J.} \bibnamefont{Rasmussen}},
  \bibnamefont{and}
  \bibinfo{author}{\bibfnamefont{D.}~\bibnamefont{Edmundson}},
  \bibinfo{journal}{J. Opt. B: Quantum Semiclass. Opt.}
  \textbf{\bibinfo{volume}{6}}, \bibinfo{pages}{S288} (\bibinfo{year}{2004}).

\bibitem[{\citenamefont{Briedis et~al.}(2005)\citenamefont{Briedis, Petersen,
  Edmundson, Krolikowski, and Bang}}]{Briedis05}
\bibinfo{author}{\bibfnamefont{D.}~\bibnamefont{Briedis}},
  \bibinfo{author}{\bibfnamefont{D.~E.} \bibnamefont{Petersen}},
  \bibinfo{author}{\bibfnamefont{D.}~\bibnamefont{Edmundson}},
  \bibinfo{author}{\bibfnamefont{W.}~\bibnamefont{Krolikowski}},
  \bibnamefont{and} \bibinfo{author}{\bibfnamefont{O.}~\bibnamefont{Bang}},
  \bibinfo{journal}{Opt. Express} \textbf{\bibinfo{volume}{13}},
  \bibinfo{pages}{435} (\bibinfo{year}{2005}).

\bibitem[{\citenamefont{Yakimenko et~al.}(2005)\citenamefont{Yakimenko,
  Zaliznyak, and Kivshar}}]{Yakimenko05}
\bibinfo{author}{\bibfnamefont{A.~I.} \bibnamefont{Yakimenko}},
  \bibinfo{author}{\bibfnamefont{Y.~A.} \bibnamefont{Zaliznyak}},
  \bibnamefont{and} \bibinfo{author}{\bibfnamefont{Y.}~\bibnamefont{Kivshar}},
  \bibinfo{journal}{\pre} \textbf{\bibinfo{volume}{71}},
  \bibinfo{pages}{065603} (\bibinfo{year}{2005}).

\bibitem[{\citenamefont{Rotschild et~al.}(2005)\citenamefont{Rotschild, Cohen,
  Manela, Segev, and Carmon}}]{Rotschild05}
\bibinfo{author}{\bibfnamefont{C.}~\bibnamefont{Rotschild}},
  \bibinfo{author}{\bibfnamefont{O.}~\bibnamefont{Cohen}},
  \bibinfo{author}{\bibfnamefont{O.}~\bibnamefont{Manela}},
  \bibinfo{author}{\bibfnamefont{M.}~\bibnamefont{Segev}}, \bibnamefont{and}
  \bibinfo{author}{\bibfnamefont{T.}~\bibnamefont{Carmon}},
  \bibinfo{journal}{\prl}  (\bibinfo{year}{2005}), \bibinfo{note}{to be
  published}.

\bibitem[{\citenamefont{Stanley and Ostrowsky}(1986)}]{FractalBook}
\bibinfo{editor}{\bibfnamefont{H.~E.} \bibnamefont{Stanley}} \bibnamefont{and}
  \bibinfo{editor}{\bibfnamefont{N.}~\bibnamefont{Ostrowsky}}, eds.,
  \emph{\bibinfo{title}{On Growth and Form}}, NATO ASI Series E: Applied
  Sciences - No. 100 (\bibinfo{publisher}{Martinus Nijhoff Publishers},
  \bibinfo{address}{Dordrecht}, \bibinfo{year}{1986}).

\bibitem[{\citenamefont{Conti et~al.}(2003)\citenamefont{Conti, Peccianti, and
  Assanto}}]{Conti03}
\bibinfo{author}{\bibfnamefont{C.}~\bibnamefont{Conti}},
  \bibinfo{author}{\bibfnamefont{M.}~\bibnamefont{Peccianti}},
  \bibnamefont{and} \bibinfo{author}{\bibfnamefont{G.}~\bibnamefont{Assanto}},
  \bibinfo{journal}{\prl} \textbf{\bibinfo{volume}{91}},
  \bibinfo{pages}{073901} (\bibinfo{year}{2003}).

\bibitem[{\citenamefont{Conti et~al.}(2004)\citenamefont{Conti, Peccianti, and
  Assanto}}]{Conti04}
\bibinfo{author}{\bibfnamefont{C.}~\bibnamefont{Conti}},
  \bibinfo{author}{\bibfnamefont{M.}~\bibnamefont{Peccianti}},
  \bibnamefont{and} \bibinfo{author}{\bibfnamefont{G.}~\bibnamefont{Assanto}},
  \bibinfo{journal}{\prl} \textbf{\bibinfo{volume}{92}},
  \bibinfo{pages}{113902} (\bibinfo{year}{2004}).

\bibitem[{\citenamefont{Rosanov et~al.}(2002)\citenamefont{Rosanov, Vladimirov,
  Skryabin, and Firth}}]{Rosanov2002}
\bibinfo{author}{\bibfnamefont{N.~N.} \bibnamefont{Rosanov}},
  \bibinfo{author}{\bibfnamefont{A.~G.} \bibnamefont{Vladimirov}},
  \bibinfo{author}{\bibfnamefont{D.~V.} \bibnamefont{Skryabin}},
  \bibnamefont{and} \bibinfo{author}{\bibfnamefont{W.~J.} \bibnamefont{Firth}},
  \bibinfo{journal}{Phys. Lett. A} \textbf{\bibinfo{volume}{293}},
  \bibinfo{pages}{45} (\bibinfo{year}{2002}).

\bibitem[{\citenamefont{Snyder and Mitchell}(1997)}]{Snyder97}
\bibinfo{author}{\bibfnamefont{A.~W.} \bibnamefont{Snyder}} \bibnamefont{and}
  \bibinfo{author}{\bibfnamefont{D.~J.} \bibnamefont{Mitchell}},
  \bibinfo{journal}{Science} \textbf{\bibinfo{volume}{276}},
  \bibinfo{pages}{1538} (\bibinfo{year}{1997}).

\bibitem[{\citenamefont{Guo et~al.}(2004)\citenamefont{Guo, Luo, Yi, Chi, and
  Xie}}]{Guo04}
\bibinfo{author}{\bibfnamefont{Q.}~\bibnamefont{Guo}},
  \bibinfo{author}{\bibfnamefont{B.}~\bibnamefont{Luo}},
  \bibinfo{author}{\bibfnamefont{F.}~\bibnamefont{Yi}},
  \bibinfo{author}{\bibfnamefont{S.}~\bibnamefont{Chi}}, \bibnamefont{and}
  \bibinfo{author}{\bibfnamefont{Y.}~\bibnamefont{Xie}},
  \bibinfo{journal}{\pre} \textbf{\bibinfo{volume}{69}},
  \bibinfo{pages}{016602} (\bibinfo{year}{2004}).

\end{thebibliography}

\end{document}